\documentclass[12pt,letterpaper]{article} 

\usepackage[margin=1in]{geometry} 
\usepackage{setspace} 
\usepackage{lineno} 

\modulolinenumbers[1] 

\usepackage{amsmath} 
\usepackage{amssymb} 
\usepackage{titlesec} 
\usepackage{graphicx} 
\usepackage{caption} 
\usepackage{enumitem} 
\usepackage{natbib} 
\usepackage[normalem]{ulem}
\usepackage{hyperref} 
\hypersetup{
    colorlinks=true,   
    linkcolor=blue,    
    citecolor=blue     
}
\usepackage{fancyhdr}       
\usepackage{multirow}
  
\usepackage{indentfirst}    
\usepackage{booktabs} 
\usepackage[table]{xcolor}
\fancypagestyle{plain}{%
    \fancyhf{}  
    \fancyfoot[R]{\footnotesize \thepage} 

}

\title{\textbf{\large Cross-Validation of Infrasound Detections and Fireball Observations from
the Global Fireball Observatory}}

\author{\normalsize Iona Clemente\textsuperscript{1,2}, Eleanor K. Sansom\textsuperscript{2}, Hadrien A. R. Devillepoix\textsuperscript{1,2},\\ \normalsize Elizabeth A. Silber\textsuperscript{3}, Svetlana B. Nikolova\textsuperscript{4}, Spiro Spiliopoulos\textsuperscript{4}, \\ \normalsize Peter Jenniskens\textsuperscript{5}, Jim Albers\textsuperscript{5}, Thomas Herring\textsuperscript{6}}

\date{} 

\begin{document}
\noindent\textit{Submitted to Journal of Geophysical Research: Planets}

\maketitle
\thispagestyle{plain}  

\small
\noindent \textsuperscript{1} Space Science and Technology Centre, School of Earth and Planetary Science, Curtin University, Perth, Australia.
\noindent \textsuperscript{2} International Centre for Radio Astronomy Research, Curtin University, Perth, Western Australia, Australia.
\noindent \textsuperscript{3} Sandia National Laboratories, Geophysics, Albuquerque, United States of America.
\noindent \textsuperscript{4} Geoscience Australia, Canberra, Australia.
\noindent \textsuperscript{5} SETI Institute, Mountain View, CA 94043, USA.
\noindent \textsuperscript{6} Jack C. Davis Observatory, Western Nevada College, Carson City, NV 89703, USA.\\

\noindent Corresponding author: iona.clemente@postgrad.curtin.edu.au\\


\section*{Abstract}
How do infrasound signals from fireballs relate to atmospheric entry conditions and the physical properties of their parents meteoroids? In this study, we investigate the detectability of infrasound generated by local-to-regional fireballs and use the associated signals to constrain meteoroid physical properties. Across Western Australia and California, the Global Fireball Observatory (GFO) recorded nearly 300 fireball events with well-constrained trajectories within 500 km of an infrasound array over an 11-year period. Of these, 20 fireballs were confidently associated with infrasound detections. For these events, we estimate energies, constrain source altitudes using atmospheric ray-tracing simulations, and investigate source mechanisms using a multi-diagnostic approach. In particular, we examine whether the recorded infrasonic signals originated from cylindrical line sources (ballistic shocks) or discrete fragmentation episodes. 
Our results show that infrasound-detected fireballs span a broad range of dynamical and physical properties. This variability indicates that infrasound detectability is governed by a combination of source characteristics, atmospheric propagation conditions, and source-receiver geometry, rather than any single entry parameter. Analysis of the fireball shock source mechanisms reveals an even contribution from both ballistic and fragmentation-related sources. These results also demonstrate the value of combining infrasound and optical observations for robust constraints on meteoroid shock generation processes.

\section*{Plain Language Summary}
When space rocks (meteoroids) blaze through the atmosphere as fireballs, they produce more than just a bright flash of light--they also generate powerful low-frequency acoustic waves called infrasound: too low for human hearing but detectable by specialised microphones. These sound waves can carry a lot of information, including whether a meteoroid broke apart, and how much energy it released. We studied nearly 300 bright fireballs recorded by camera networks in Western Australia and California and found infrasound for 20 of them. Combining camera and acoustic data let us pinpoint where each signal originated, and whether it came from a sudden fragmentation event or from the meteoroid's passage through the atmosphere at hypersonic speeds. In this dataset, we found that both mechanisms were equally common, and whether a fireball was detected at all depends on a complex mix of factors, not just brightness and distance. 

\section{Introduction}
Infrasound refers to low-frequency acoustic signals in the range from 0.01 Hz, the natural oscillation frequency of the atmosphere, to 20 Hz, the lower limit of human hearing. Infrasound is particularly suitable for fireball detection as low-frequency acoustic waves experience very little attenuation, approximately proportional to the square of the frequency \citep{bass1972atmospheric, sutherland2004atmospheric}. A comprehensive statistical analysis of infrasound observations from small meteoroids was previously conducted by \citet{silber2014optical}. Since then, most studies of fireball-generated infrasound have focused on datasets dominated by large fireballs (e.g., CNEOS-class events; \citet{pilger2020global,silber2025investigating}). Studies targeting smaller fireballs have generally relied on dense  seismo-acoustic networks (e.g., \citet{scamfer2026multi}). This raises the question of how much information about local-to-regional fireballs can be extracted from single-station infrasound detections.\\

The fireball phenomenon results from the rapid deceleration of the passage of a meteoroid through the Earth's atmosphere. In most cases, the object breaks up and entirely vaporizes \citep{ceplecha2005fragmentation}. However, sometimes it may survive longer and, in some cases, even drop a meteorite on the ground. When bodies are large enough (greater than ~1-10 centimeters in diameter), meteoroids can generate shock waves \citep{bronshten1964problems, revelle1976meteor, revelle2008acoustic, silber2018infrasound, silber2018physics} recordable by ground-based seismo-acoustic sensors. 
 
These shock waves may arise from different processes occurring at various stages along the fireball trajectory. (1) The first mechanism is the hypersonic passage of the object through the atmosphere that creates a conical bow shock called a Mach cone \citep{revelle1976meteor}. As meteoroids enter Earth's atmosphere at velocities ranging from 11 to 73 km/s (equivalent to Mach number 35-240: \cite{ceplecha1998meteor, murad2002meteors}), their velocity overtakes the speed of sound. Therefore, the Mach cone angle being extremely small allows the cone's wavefront to be approximated as a cylinder \citep{tsikulin1970shock, revelle1976meteor}. Such a propagation of the shock wave is assumed to be perpendicular to the meteoroid's trajectory. It also induces that the direct arrival from this ballistic trajectory will match the arrival time of a shock wave coming from the shortest distance to the trajectory. (2) The second process of generating shock waves is a fragmentation event \citep{revelle1976meteor, ceplecha1998meteor,silber2018physics}. Such an event occurs when the force exerted by the motion of the object through the atmosphere (ram pressure) on its surface exceeds its compression strength, causing it to break into smaller fragments \citep{cevolani1994explosion,stevanovic2017bolide}. (3) Another mechanism that can produce shock waves is a catastrophic final airburst, which corresponds to the complete disintegration of the meteoroid into fine dust \citep{klekociuk2005meteoritic}. Both airbursts and fragmentation events generate shock waves propagating quasi-spherically \citep{edwards2008seismic}.\\

\begin{figure}[!h]
    \centering
    \includegraphics[width=0.75\textwidth]{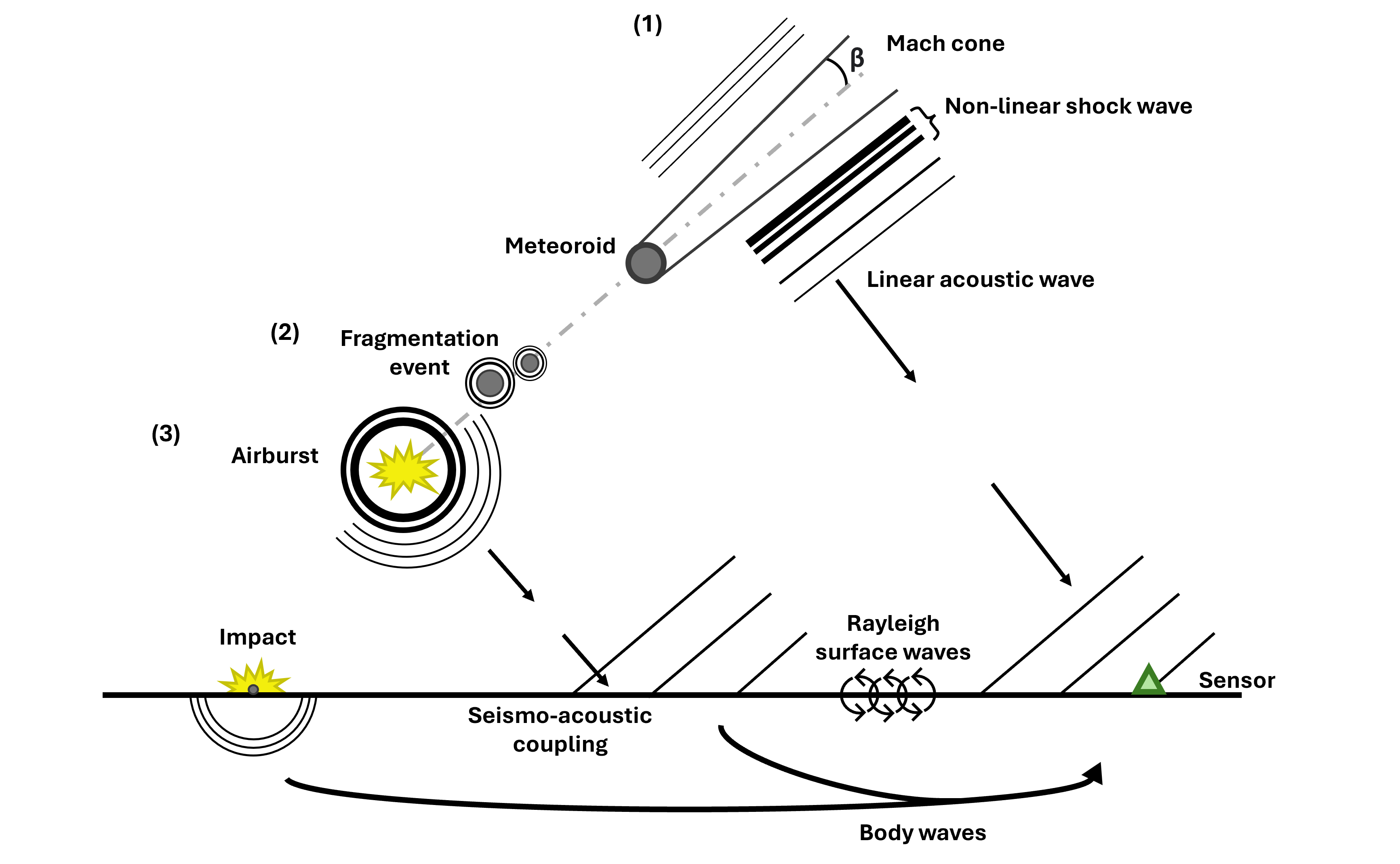}
    \caption{Different types of waves and their associated generating events that may arise during a fireball event: (1) Mach cone generating shock waves that travel almost perpendicular to the fireball's trajectory, (2) Fragmentation event generating shock wave that propagate omnidirectionally, (3) Final catastrophic airbust producing shock wave travelling omnidirectionally. In the case of the meteoroid impacting the ground, it is also possible to generate seismic surface and body waves (extremely rare). Figure redrawn from \cite{edwards2008seismic} and \cite{neidhart2021statistical}.}
    \label{fig:SW_gen}
\end{figure}

The Desert Fireball Network (DFN) is a network of optical observatories that has been monitoring the Australian night sky for over eleven years. Its primary goal is to detect fireballs and determine their trajectories and orbits to enable meteorites recoveries \citep{devillepoix2018dingle, devillepoix2019observation}. The DFN has expanded beyond Australia, and observatories have been deployed on several continents, including Europe, Middle East and America, as part of the international Global Fireball Observatory (GFO) collaboration. Since 2014, DFN and GFO cameras have recorded over 2000 global events from objects $>$1 cm in diameter up to May 2025. \\

The Preparatory Commission for the Comprehensive Nuclear-Test-Ban Treaty Organization (CTBTO PrepCom; \cite{Pichon2019infrasound}) was established to monitor and detect nuclear test explosions. To achieve this, it operates a global network of sensors, including infrasound arrays, as part of the International Monitoring System (IMS; \cite{christie2009ims}).
In this study, we investigate a subset of bright fireballs detected near a single IMS infrasound station to identify meteor-generated infrasound from regional events. Using high-fidelity trajectories and multi-modal observations, we estimate fireball energies, constrain shock source altitudes, and assess the detection capability of infrasound for regional events. We also investigate the origin of the recorded signals by distinguishing between cylindrical line-source and fragmentation-generated shock mechanisms. \\

\section{Data and Methods}

\subsection{Data and Processing}

For the list of fireballs to investigate, we use the confirmed multi-station detections by the Global Fireball Observatory (GFO).
These sensors enable the reconstruction of the meteoroid's atmospheric path to typically $\sim$100 m in space, and $\sim$0.1 s in absolute time \citep{howie2017build, sansom2019determining}.
Infrasound signals were recovered from two infrasound arrays from the IMS operated by CTBTO -- IS04 and IS57. Each IMS’s infrasound array consists of several microbarometers (sensitive microphones) recording low frequency (1-20 Hz) atmospheric pressure variations generated by explosions and other energetic events. These arrays are commonly used for beamforming analyses to estimate the back-azimuth and trace velocity of incoming signals \citep{bishop2020improved, cansi2008infrasound}.\\

GFO fireballs were selected based on their proximity to the IS04 and IS57 infrasound stations, located in Western Australia and California (US), respectively.
Initially, events were included if any portion of their trajectory (either the start or end point) lay within a 200 km radius of an array. Following this selection, infrasound records for the corresponding dates were obtained from CTBTO and manually processed and checked for plausible signals. As detections were yielding promising results, the search area was expanded to a radius of 500 km around the infrasound stations. \\

For each event, we examined the IMS data over the entire day of occurrence of the fireball to assess the general background noise level. The data were first processed using the Progressive Multi-Channel Correlation (PMCC) method \citep{brachet2009monitoring} to identify potential shock wave signatures associated with the event. A signal was considered a candidate fireball-generated detection when both its arrival time and back-azimuth were consistent with the theoretical arrival time and back-azimuth predicted from the endpoint of the fireball bright trajectory. 
Once candidate signals were confirmed, raw infrasound data were further processed in Python \citep{rossum2009python} using the seismo-acoustic library ObsPy \citep{beyreuther2010obspy, krischer2015obspy}, with the goal of computing meteoroid energy deposition estimates.

\noindent Instruments responses were first removed using the \textit{obspy.signal} module. A power spectral density (PSD) analysis was then performed to determine appropriate corner frequencies for filtering. For each event, traces were trimmed to a time window spanning from –20 min to +40 min relative to the signal onset identified by the PMCC analysis. The PSD was computed and plotted for (i) a signal window centred on the detected arrival and (ii) a noise window defined from intervals before and after the event. \\

\noindent Cutoff frequencies were selected at the intersection points where the signal PSD exceeds the noise PSD and where it subsequently drops below it (see \cite{ens2012infrasound, silber2024ground}). The traces were then bandpass filtered using the previously identified PSD corner frequencies. Finally, for each event, traces from all elements of the infrasound array were stacked to enhance the signal-to-noise ratio, excluding those identified as excessively noisy. \\

\begin{figure}[!h]
    \centering
    \includegraphics[width=0.99\textwidth]{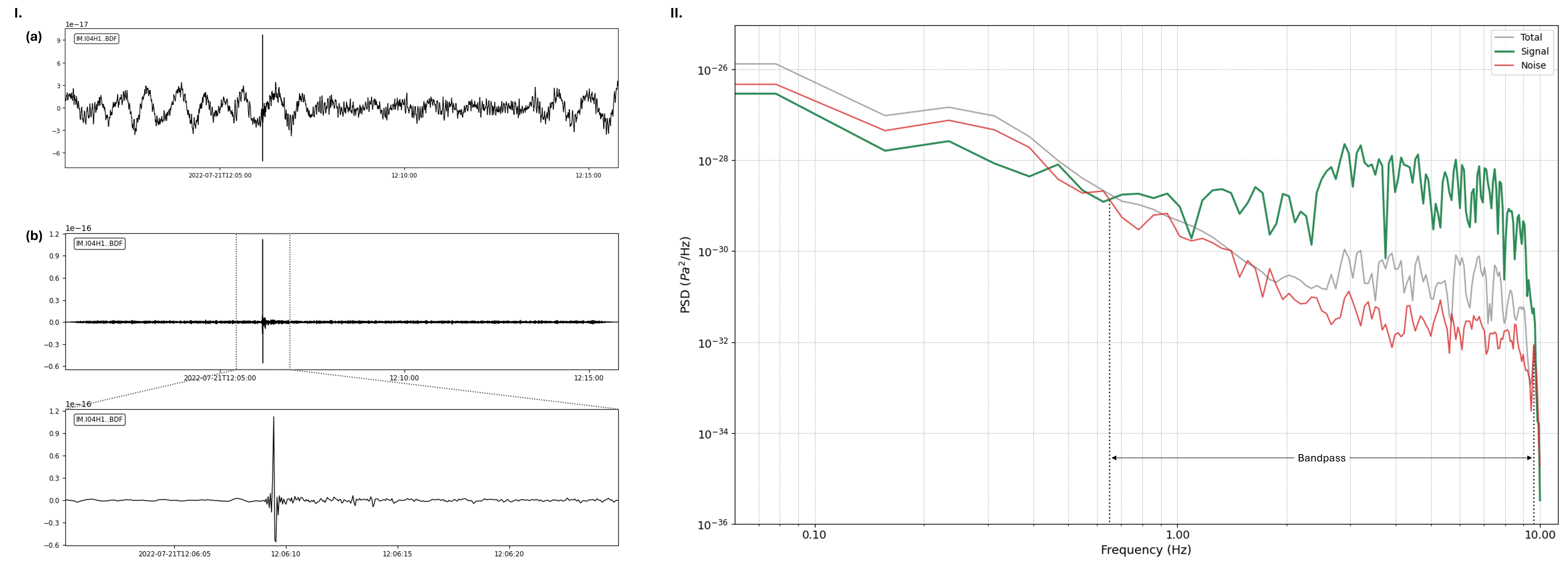}
    \caption{Signal processing steps illustrated using the example of fireball event DN220721\_01. Panel I.a. shows the raw signal, while Panel I.b. presents the filtered signal. The lower plot provides a zoomed-in view of the filtered waveform. Panel II. shows the frequency content of the fireball signal. The green curve represents the power spectral density (PSD) of the signal only, obtained by subtracting the noise PSD (red curve) from the total signal PSD (grey curve). The dashed black lines indicate the corner frequencies of the bandpass filter (here 0.65-9.5 Hz), defined as the frequencies on either side of the PSD maximum where the signal PSD decreases to the level of the noise PSD.}
    \label{fig:psd}
\end{figure}

The final step involved determining the signal period from the zero-crossing method \citep{revelle1997historical}, specifically the four zero-crossing (4-ZC) approach described in \cite{silber2025multiparameter}, Appendix. This period is hereafter referred to as the zero-crossing period, $T_{ZC}$.
Using these periods estimates, we can then compute the corresponding energy estimates for the fireball events ($E_{ZC}$ in Tab. \ref{table:EdepMassRad}). \\

\subsection{Energy estimates}\label{sec:methods_energy}

Event energy estimates were obtained using the refined empirical period-yield relation proposed by \cite{silber2025multiparameter}, which links the energy deposition of an event to the period of its infrasonic signal. This formulation builds on empirical period-yield  relations derived from nuclear test data (AFTAC (U.S. Air Force Technical Applications Center)) and adapted for bolides by ReVelle (1997). \cite{ens2012infrasound} (Section 2) demonstrated that these explosion-based period-yield relations hold remarkably well for bolides and provide useful empirical expressions specific to global bolide datasets. Recently, as the global bolide dataset has expanded, the period-yield relations were further refined (Silber et al. 2025). 
For single-station recordings, as in our case, the period-yield relation is expressed as follows:
\begin{equation}
    \log(E) = 3.30\pm 0.14 \log(P) - 1.89\pm0.10 
    \label{eq:edep_silber}
\end{equation}
where \textit{E} denotes the fireball energy deposition in kt of TNT equivalent and \textit{P} the signal period in seconds. All logarithms are base 10. In this study, energy estimates were calculated using $T_{ZC}$ as input periods.\\

\subsection{Shock Source Altitude estimates}

Estimates of shock source altitudes were obtained using the ray-tracing package \textit{infraGA} \citep{blom2024}, which was used to simulate all possible acoustic propagation paths from candidate sources along the trajectory to the receiver (infrasound sensor), using real atmospheric profiles from Ground-to-Space (G2S) models \citep{drob2003global, hetzer2024ncpag2s}. For each event, the atmospheric profile was extracted at the time of the fireball and at the end point of the bright flight.\\

For each fireball event, source candidates were defined along the trajectory from initial to final altitude at 0.5 km intervals. For each candidate, position, emission time, and source-receiver azimuth were interpolated from the trajectory endpoints. Rays were launched using infraGA in 3D propagation mode. Azimuths were sampled within $\pm10^\circ$ of the interpolated source-receiver azimuth at $1^\circ$ intervals, while launch inclinations spanned $-89^\circ$ to $+89^\circ$ with a $1^\circ$ step. Ground reflections were permitted.\\

\begin{figure}[!h]
    \centering
    \includegraphics[width=0.95\textwidth]{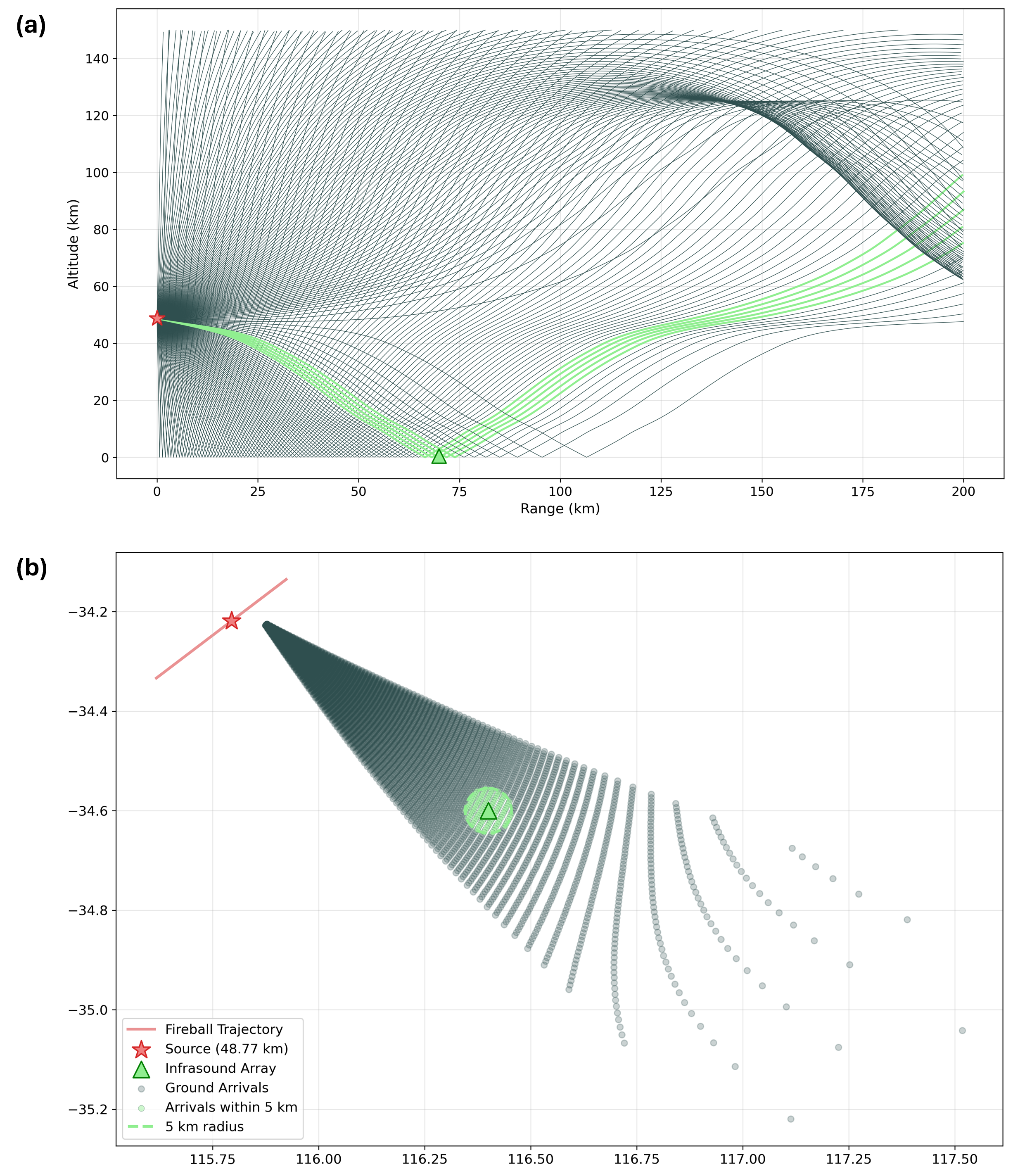}
    \caption{Example of an infraGA ray-tracing simulation of shock wave propagation for the DN220721\_01 fireball event. The source height of 48.77 km corresponds to the mean source altitude estimate for this event (red star). (a) Altitude-range profile of ray propagation over a 200 km range. Green-highlighted paths indicate rays that reach the ground within a 5 km radius of the receiver. (b) Map view of ground-level ray arrivals. Green markers denote arrivals occurring within a 5 km radius of the infrasound array, used to constrain the source altitude window.}
    \label{fig:infraga2d3d}
\end{figure}

\begin{figure}[!h]
    \centering
    \includegraphics[width=0.7\textwidth]{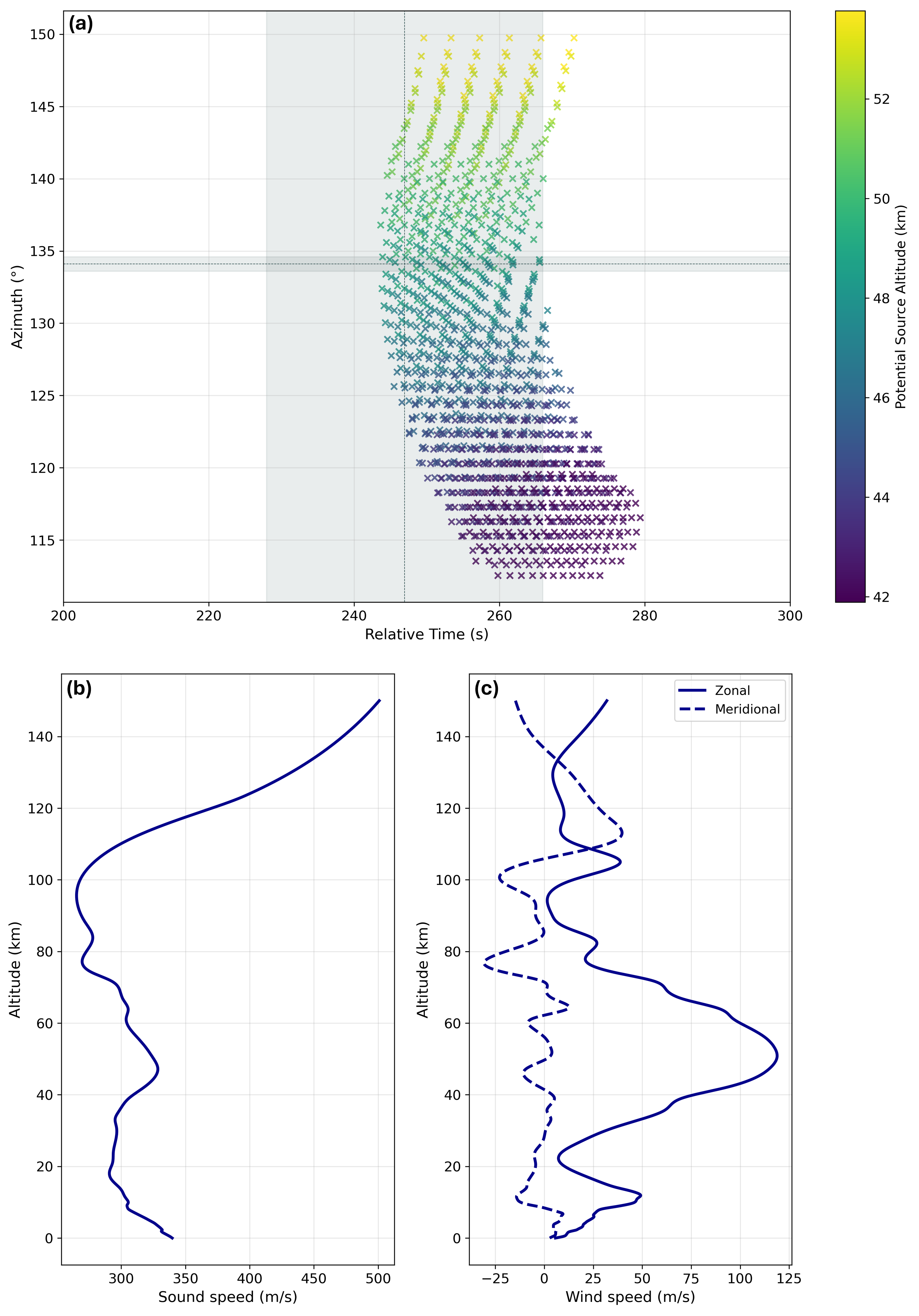}
    \caption{(a) Illustration of the shock wave source altitude determination method for the DN220721\_01 fireball event. Each coloured cross represents a potential source altitude along the trajectory. The signal azimuth is plotted as a function of relative time since the start of the bright flight. The solid lines indicate the observed azimuth and arrival time of the signal, while the surrounding grey shading represents their associated uncertainties. The overlap between modelled points and observed constraints identifies the most likely source altitude of the shock. (b) Thermodynamic sound speed profile over the DN220721\_01 fireball site. (c) Wind profiles showing the zonal (solid line) and meridional (dashed line) wind components.}
    \label{fig:SourceAlt}
\end{figure}

A pre-selection was first applied to retain only arrivals located within a 5 km radius of the infrasound sensor. If no arrivals were found, the search radius was extended by 5 km increments. Among these pre-selected rays, those matching both the observed signal arrival time and azimuth (within their respective uncertainties; see Fig. \ref{fig:SourceAlt}) were selected, and the corresponding source altitude candidates were identified. When multiple solutions were obtained, only the subset of candidate source altitudes forming a continuous range was retained, excluding isolated values. The mean value, associated uncertainty, and standard deviation were then computed from this subset.\\

\subsection{Shock Source Mechanisms}

\begin{figure}[!h]
    \centering
    \includegraphics[width=0.5\textwidth]{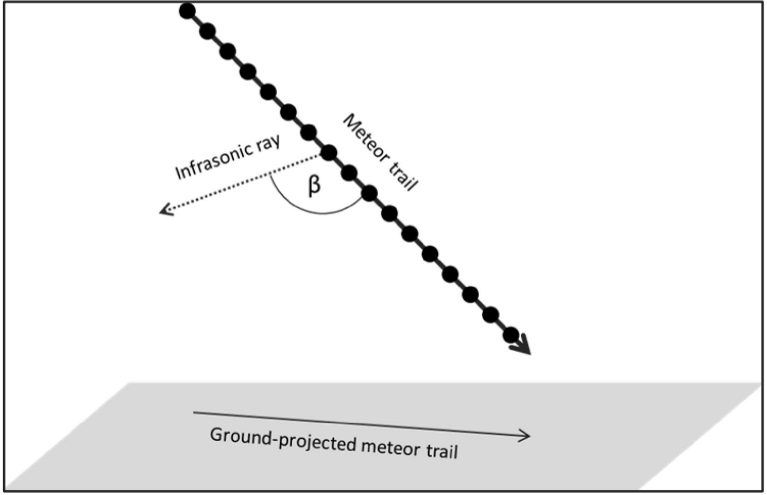}
    \caption{Ray-tracing modelling procedure to retrieve fireball shock wave source altitude estimate. Black circles represent distinct candidate source points along the fireball trajectory. Infrasonic rays are launched from each point toward the infrasound sensor (only one ray is shown here for simplicity). The deviation angle $\beta$ corresponds to the 3D angle between the ray direction and the meteoroid velocity vector. Figure from \cite{silber2014optical}.}
    \label{fig:raydev}
\end{figure}

A shock wave generated by a cylindrical line source (ballistic trajectory) is expected to propagate approximately perpendicular to the fireball trajectory, whereas a shock wave produced by a fragmentation event has an omnidirectional propagation. Based on this distinction, we first investigated the source type of the recorded shock waves by computing the ray deviation from the trajectory (see Fig. \ref{fig:raydev}). In this study, this deviation is defined as the angle the fireball trajectory and the direction of the launched ray at the estimated mean source altitude. \\

\noindent If this angle falls within the $90\pm25^\circ$, the source is interpreted as consistent with a ballistic trajectory origin \citep{revelle1976meteor, brown2007acoustic}. Conversely, if the angle lies outside the $90\pm25^\circ$ range, the shock wave is interpreted as originating from one (or more) fragmentation episode \citep{revelle1976meteor, brown2007acoustic}. However, a fragmentation event may also produce a shock wave that appears to propagate perpendicular to the trajectory due to the specific geometry between the trajectory and the sensor. In such cases, the source type cannot be unambiguously determined from the ray deviation alone. \\

We therefore complemented this analysis by examining both the infrasound waveforms and optical observations, including light curves, to identify potential signatures of fragmentation. 
Light curves can reveal fragmentation episode(s) through distinct peak(s) in the fireball’s intensity profile. In addition, the infrasound signals were analysed based on waveform characteristics described in \cite{silber2024perspectives} (Fig. 7). This study showed that infrasound signals with an N-wave or double N-wave morphology, sometimes accompanied by a coda, are typically indicative of a cylindrical line source (i.e., ballistic origin). In contrast, signals composed of multiple arrivals–such as multiple N-waves or multiple diffuse arrivals–are generally associated with fragmentation events. A single diffuse arrival may also indicate a far-field detection.\\

Another method for assessing fragmentation probability, based on the fireball trajectory dynamics, was proposed by \citet{sansom2016filtering}. It leverages two Bayesian filters within an interacting multiple model (IMM) framework to identify two trajectory states: an ablation-only mode and a high mass loss one, likely associated with fragmentation. This allows gross fragmentation behaviour to be modelled statistically. We ran IMM trajectory models to our fireball events following the method of \citet{sansom2016filtering}, including the same mode switching parameters given in Eq. 22 of their work, and discuss where the 'fragmentation' mode is likely dominant.


\section{Results}

\subsection{Detectability}

\begin{figure}[!h]
    \centering
    \includegraphics[width=0.99\textwidth]{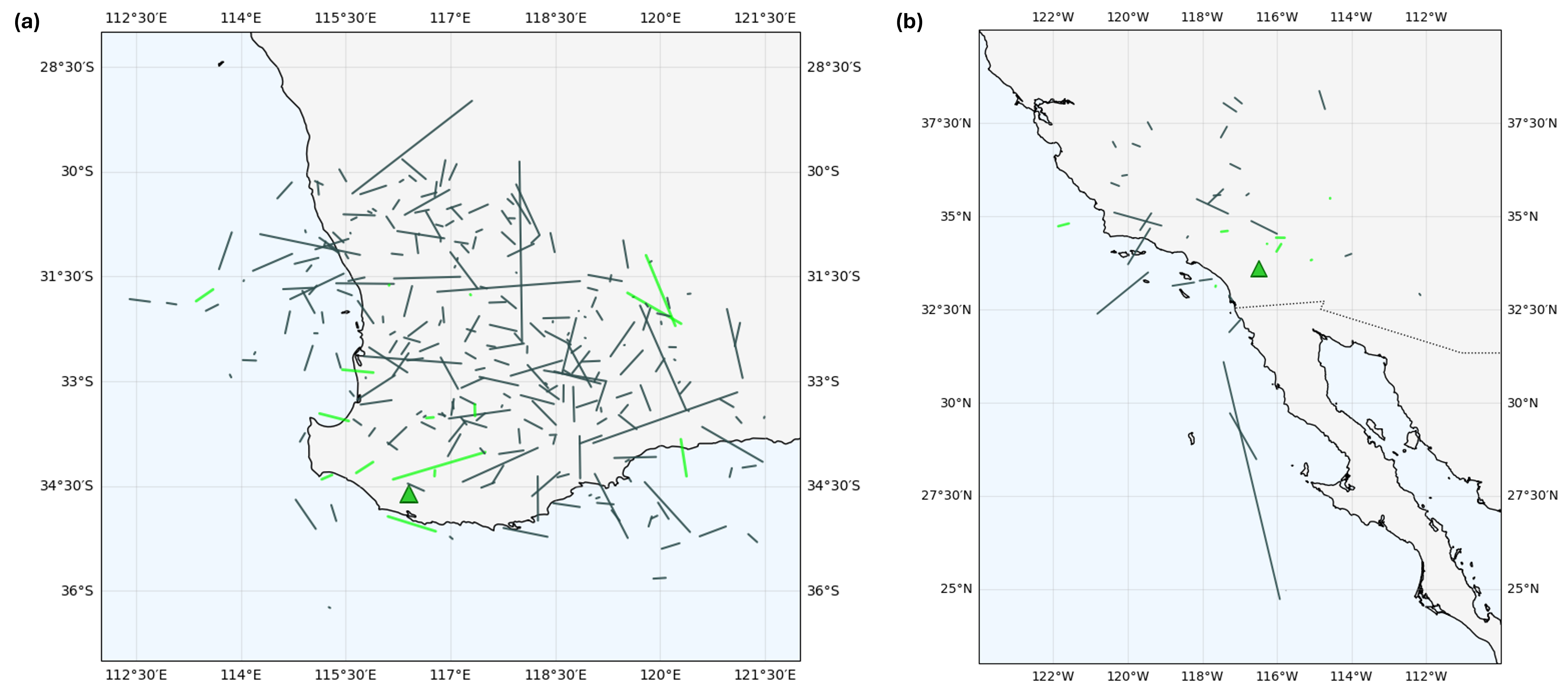}
    \caption{Maps showing fireballs observed by DFN and GFO cameras in the vicinity of two infrasound arrays: (a) southern Western Australia near IS04 and (b) southern California (USA) near IS57. Green triangles mark the infrasound arrays locations. Fireball trajectories are shown in green for events with infrasound detections and in grey for non-detections.}
    \label{fig:map}
\end{figure}

Of the 296 fireballs analysed, 260 occurred near the IS04 infrasound station and 36 near IS57 (the disparity is mostly due to differences between the camera  networks' spatial arrangement and effective operational time). Using the 200 km limit, 15 fireballs were linked to infrasound detections (9 in Western Australia and 6 in California). Extending the search radius to 500 km added a further 6 events in Western Australia and 2 in California (see Fig. \ref{fig:map}), increased the total number of detections to 23. These events occurred at 3D shortest distances between the sensor and the fireball trajectory ranging from approximately 66 to 491 km. Detailed information on the fireball bright flight start times and sensor distances is provided in Table \ref{table:GeneralInfoFB} (Appendix). \\

\begin{table*}[!h]
    \caption{Summary of energy estimates for the 23 fireballs with detected infrasound signals. $f_D$ denotes the dominant frequency of the signal. The \textit{Bandpass} column corresponds to the frequencies used for bandpass filtering during signal processing. The \textit{Range} column corresponds to the shortest distance between the fireball trajectory and the infrasound array. Gray-highlighted events correspond to fireball located more than 200 km from an infrasound array.}
    \label{table:EdepMassRad}
    \centering
    \resizebox{0.8\textwidth}{!}{
        \begin{tabular}{c c c c c c}
        \hline
        Fireball ID & Sensor & $f_D$ (Hz) &  Bandpass (Hz) & $E_{ZC}$ (kt) & Fireball-Sensor Range (km)\\
        \hline\hline
\rowcolor{gray!20}DN161030\_04 & IS04 & 5.39 & 2.90 - 6.89 & 0.000157 & 367.3 \\
DN171009\_01 & IS04 & 2.11 & 1.27 - 7.92 & 3.933E-05 & 167.4 \\
DN180709\_01 & IS04 & 5.63 & 1.11 - 9.50 & 4.476E-05 & 198.0 \\
\rowcolor{gray!20}DN180828\_01 & IS04 & 1.66 & 0.50 - 5.27 & 0.017992 & 327.3 \\
DN190824\_02 & IS57 & - & - & - & 142.7 \\
DN191001\_02 & IS57 & - & - & - & 98.7 \\
DN191002\_01 & IS04 & 6.60 & 0.79 - 9.50 & 0.000328 & 65.6 \\
\rowcolor{gray!20}DN191122\_01 & IS04 & 2.19 & 0.44 - 3.23 & 0.001496 & 433.1 \\
DN200512\_04 & IS04 & 5.23 & 0.94 - 7.20 & 0.001459 & 79.4 \\
DN200710\_01 & IS04 & 2.81 & 0.92 - 8.70 & 0.005184 & 163.0 \\
DN201213\_04 & IS57 & 8.13 & 3.1 - 8.98 & 7.291E-05 & 89.7 \\
DN210113\_02 & IS57 & 1.80 & 0.55 - 9.14 & 0.000831 & 150.6 \\
\rowcolor{gray!20}DN210211\_01 & IS04 & 2.11 & 1.06 - 2.88 & 0.001372 & 334.6 \\
DN210319\_01 & IS04 & 3.13 & 0.64 - 9.50 & 0.000572 & 87.4 \\
\rowcolor{gray!20}DN210418\_02 & IS04 & 2.42 & 1.13 - 9.37 & 0.000639 & 418.2\\
\rowcolor{gray!20}DN210904\_01 & IS57 & 6.41 & 1.74 - 9.50 & 1.727E-05 & 273.4\\
DN211013\_03 & IS57 & 5.94 & 0.91 - 9.50 & 0.000124 & 118.0 \\
DN211106\_01 & IS04 & 6.56 & 5.84 - 8.58 & 1.719E-05 & 131.3 \\
DN220721\_01 & IS04 & 3.91 & 0.65 - 9.50 & 0.000185 & 85.2 \\
\rowcolor{gray!20}DN230302\_02 & IS57 & 5.31 & 3.17 - 9.50 & 6.712E-05 & 491.2 \\
DN230629\_02 & IS04 & 6.41 & 1.22 - 9.32 & 6.020E-05 & 120.9 \\
DN231006\_01 & IS57 & 4.84 & 0.83 - 9.50 & 0.000172 & 118.5 \\
\rowcolor{gray!20}DN250510\_01 & IS04 & 2.30 & 0.78 - 3.12 & 0.010417 & 443.2\\
        \hline
        \end{tabular}
    }
\end{table*}

Among the 23 events with detected infrasound signals, energy deposition estimates were successfully computed for 21 events using the method of \ref{sec:methods_energy}. For the remaining two events, despite clear signals, issues during the instrument response removal prevented the calculation of their absolute energy estimates (Table \ref{table:EdepMassRad}). They were however retained in the dataset and considered for the further analyses.

\begin{figure}[!h]
    \centering
    \includegraphics[width=0.99\textwidth]{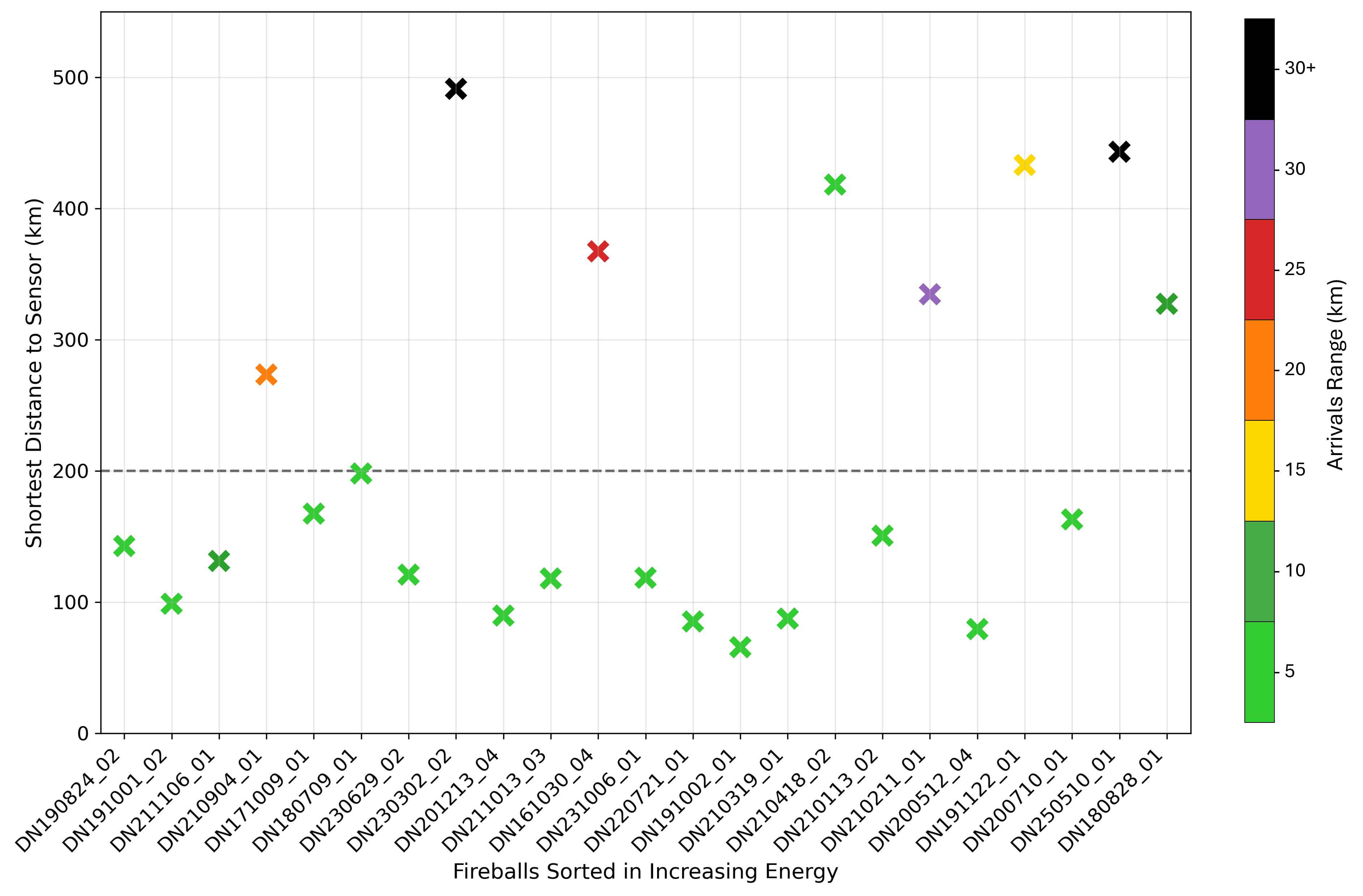}
    \caption{Fireball events ordered by increasing energy deposition as a function of the shortest distance between the fireball trajectory and the infrasound sensor (km). The colorbar indicates the arrival-distance tolerance range (km) within which the infraGA-simulated rays were selected during the shock source altitude window estimation process. Events for which energy estimates could not be determined (DN190824\_02 and DN191001\_02) are placed at the beginning of the sequence. Fireball events marked with a black cross indicate cases where the ray-tracing simulations could not reproduce the observed arrivals, even when the arrival range tolerance was increased beyond 30 km.}
    \label{fig:infraga_arriv_range}
\end{figure}

The computed energies range from $1.719\times10^{-5}$ to $1.799\times10^{-2}$ kt TNT equivalent. We expected the more distant events ($>$200~ km) to be associated with the highest-energy fireballs, however, several lower-energy events were also detected at large distances. In particular, three events (DN161030\_04, DN210904\_01 and DN230302\_02) with estimated energies below 0.2 t TNT eq. were detected at distances ranging from over 250 km to nearly 500 km from the infrasound array (Fig. \ref{fig:infraga_arriv_range}). Despite the consistency in arrival time and back-azimuth, this raised some uncertainty regarding the validity of these three detections being associated with our fireball events. This is further explored in the following analysis.\\

Within the 200 km radius limit, ray-tracing modelling allowed us to estimate shock source altitudes for all 15 fireballs using pre-selected arrivals located within a 5-10 km radius of the infrasound array (Fig. \ref{fig:infraga_arriv_range}). This spatial constraint was applied to ensure that only arrivals representative of local propagation conditions and short source-to-receiver distances were considered. Ray tracing also indicated that all of these detected signals were direct arrivals. \\

\noindent For the eight additional fireballs identified in the extended search area beyond 200~km, obtaining source altitudes using ray tracing was more challenging. We were only able to determine windows for source altitudes for two of them while maintaining the 5-10 km constraint: DN210418\_02 and DN180828\_01. Notably, DN180828\_01 corresponds to the highest estimated energy deposition in our dataset. We are therefore confident in these two event signals. 
For the six remaining fireballs located beyond 200~km, source altitude windows were determined for four events only after further extending the acceptable arrival range (from 15 up to 30 km; see Fig. \ref{fig:infraga_arriv_range}). For the two other events (DN230302\_02 and DN250510\_01; black crosses in Fig. \ref{fig:infraga_arriv_range}) we were unable to match the observed arrivals with the modelled ones within a 30~km radius of the receiver. These events also correspond to the two most distant detections, at 491 km and 443 km, respectively. Ray tracing results also indicated a change in propagation regime for these more distant events, with arrivals involving reflected and/or ducted phases rather than purely direct arrivals. This likely indicates a stronger dependence on the atmospheric profiles adopted for more distant events.\\

\noindent Overall, for the events beyond 200~km from the infrasound stations, five signals are considered as reliably originating from the corresponding fireball. Although the range of modelled arrivals needed to be increased for some, these are higher energy events. This includes event DN250510\_01 where several meteorite fragments were recovered by the DFN team (see Clemente et al., in prep.). Events DN210904\_01, DN230302\_02 and DN161030\_04 were already considered uncertain due to their large source-to-receiver distances despite their low estimated energies. Additionally, the arrival range had to be extended beyond a 15~km radius from the receiver, this difficulty in matching observed and modelled arrivals means the attribution of the detected infrasound signals to the corresponding fireball events is considered more unlikely. These events were therefore excluded from the rest of this study, and our statistics as confirmed infrasound detections.\\

In summary, energy estimates were obtained for 18 of the 20 fireballs with confirmed infrasound detections, yielding values ranging from $1.719 \times 10^{-5}$ to $1.799 \times 10^{-2}$ kt TNT equivalent. Shock source altitudes were determined for 19 of the 20 events, spanning 41-87 km. Associated uncertainties, where estimable, range up to 15.5 km, although most events (15/19) exhibit uncertainties below 5 km. Individual mean source altitudes, associated uncertainties, and standard deviations are reported in Table \ref{table:SourceAlt_NewMass}.

\begin{table*}[!h]
    \caption{Summary of shock wave source altitudes and associated uncertainties, and standard deviation for the 20 fireball events with detected infrasound signals.}
    \label{table:SourceAlt_NewMass}
    \centering
    \resizebox{0.8\textwidth}{!}{
        \begin{tabular}{c c c c c}
        \hline
        Fireball ID & Sensor & Source Altitude (km) & Std (km) & Range Arrivals (km) \\
        \hline\hline
DN171009\_01 & IS04 & $80.23\pm3.5$ & 2.16 & 5 \\
DN180709\_01 & IS04 & 42.86 & -  & 5\\
DN180828\_01 & IS04 & $79.65\pm0.25$ & 0.25 & 10 \\
DN190824\_02 & IS57 & $52.64\pm7.25$ & 4.33 & 5 \\
DN191001\_02 & IS57 & $59.38\pm15.50$ & 9.09 & 5 \\
DN191002\_01 & IS04 & $48.21\pm10.25$ & 6.06  & 5\\
DN191122\_01 & IS04 & $53.67\pm0.50$ & 0.41 & 15 \\
DN200512\_04 & IS04 & $71.86\pm0.50$ & 0.41 & 5 \\
DN200710\_01 & IS04 & $87.12\pm0.25$ & 0.25 & 5 \\
DN201213\_04 & IS57 & $49.38\pm6.50$ & 3.89 & 5 \\
DN210113\_02 & IS57 & $56.52\pm4.25$ & 2.59 & 5 \\
DN210211\_01 & IS04 & 72.27 & - & 30 \\
DN210319\_01 & IS04 & $66.35\pm3.25$ & 2.02 & 5 \\
DN210418\_02 & IS04 & $64.83\pm0.50$ & 0.50 & 5 \\
DN211013\_03 & IS57 & $41.12\pm2.25$ & 1.44 & 5 \\
DN211106\_01 & IS04 & $52.39\pm1.75$ & 1.15 & 10 \\
DN220721\_01 & IS04 & $48.77\pm1.50$ & 1.00 & 5 \\
DN230629\_02 & IS04 & $60.41\pm4.50$ & 2.17 & 5 \\
DN231006\_01 & IS57 & $59.15\pm0.75$ & 0.56 & 5 \\
DN250510\_01 & IS04 & - & - & - \\
        \hline
        \end{tabular}
    }
\end{table*}

For all events with source-to-receiver distances below 200 km, direct arrivals were observed, whereas more distant detections involved ground and/or atmospheric layer reflections. No preferential shock source region was identified along the fireball trajectories, as inferred source altitudes are distributed throughout the luminous paths.\\

\begin{figure}[!h]
    \centering
    \includegraphics[width=0.99\textwidth]{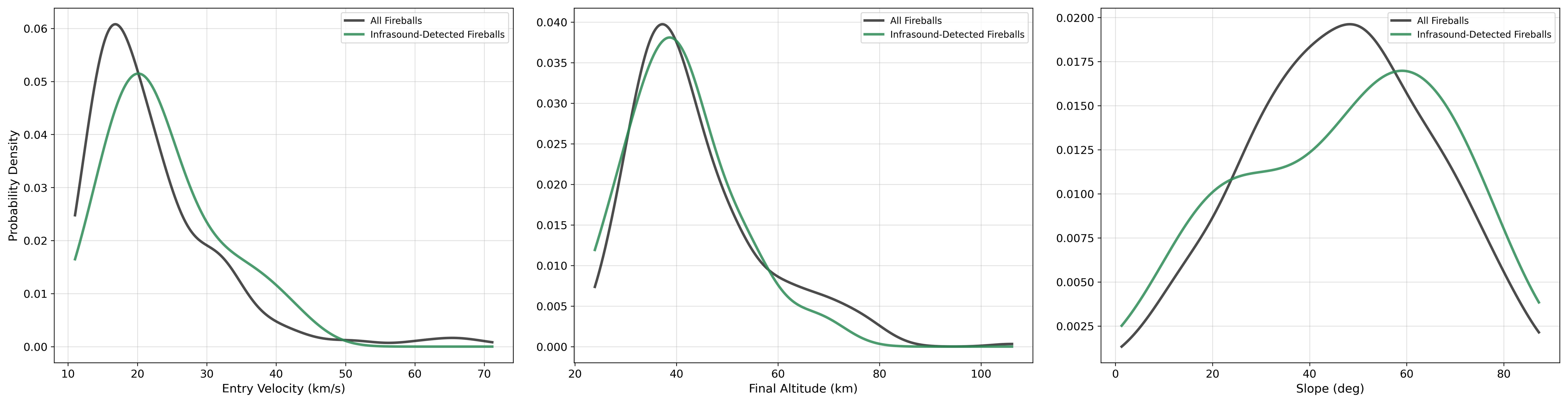}
    \caption{Kernel Density Estimation (KDE) analysis comparing the complete fireball dataset (black curves) with the subset of 20 fireball events associated with confirmed infrasound detections (green curves), shown for different trajectory parameters. From left to right: probability density distributions of fireball entry velocity (km/s), final altitude (km), and entry angle (degrees).}
    \label{fig:kde}
\end{figure}

Using the refined subset of confirmed infrasound detections, we also performed a Kernel Density Estimation (KDE) analysis to compare the distributions of key parameters--entry velocity, final altitude, and entry angle--with those of the complete fireball dataset. The objective was to determine whether detected events preferentially cluster within specific parameter ranges. Aside from a slight shift toward higher entry velocities for infrasound-detected events, no significant differences are observed between the two distributions (Fig. \ref{fig:kde}). A complementary principal component analysis (PCA) led to the same conclusion, showing no distinct clustering of infrasound-detected events relative to the full fireball population.

\subsection{Shock Source Mechanisms}

\subsubsection*{Ray Deviation}
In order to assess the shock source mechanism, we first determine ray deviation \citep{silber2014optical}. Among our final subset of 20 fireballs, the ray deviation analysis was performed for the 19 events for which a shock source altitude window could be estimated. A ray deviation angle outside the $90\pm25^\circ$ range would suggest a fragmentation origin. Only one fireball (DN201213\_04) falls outside this range. For the remaining 18 events, whose ray deviation angles fall within the $90\pm25^\circ$ interval, the shock source mechanism could not be uniquely constrained, as both ballistic and fragmentation origins remain possible (Table \ref{table:Fragm?}, Appendix).

\subsubsection*{Signal Morphology}
\noindent The next step for determining shock source mechanism is to assess the morphology of the infrasound signals. We find that 10 fireball shock sources are consistent with a ballistic origin, with waveforms exhibiting an N-wave, or in some cases a double or triple N-wave (e.g. Fig. \ref{fig:shockmechanism}a, left; DN210319\_01 shows a clear N-wave). In contrast, 10 events are more likely associated with one or several fragmentation episodes, characterized by more diffuse waveforms and/or multiple arrivals (e.g. Fig. \ref{fig:shockmechanism}a, right; DN180709\_01 illustrates a diffuse detection).\\

\subsubsection*{Light curves}
\noindent Of the 20 fireball events analysed, light curves were available for 17. However, the current DFN light curves still have some limitations associated with the processing pipeline. In particular, they provide apparent rather than absolute magnitudes, and some are saturated. These limitations restrict the detailed identification of fragmentation features. Nevertheless, some light curves are consistent with results from our analysis and are shown in Figure \ref{fig:shockmechanism} (the others can be found in Fig. \ref{fig:LCs}, Appendix). For example, the light curve of DN191122\_01 shows strong evidence of fragmentation, consistent with the source altitude window inferred from the ray-tracing simulations. Another interesting example is DN250510\_01, whose light curve likewise exhibits fragmentation behaviour. This interpretation is further supported by the recovery of three meteorites from this event (Clemente et al., 2026, in prep.), providing independent confirmation of the fragmentation scenario.\\

\subsubsection*{IMM}
Meteoroid flight dynamics are modelled using a set of single-body aerodynamic equations describing the evolution of position, velocity, and mass through the atmosphere. As per \citet{sansom2016filtering}, two models run simultaneously: Model 1 carries a low mass-loss uncertainty, representative of steady ablation, while Model 2 carries a significantly higher mass-loss uncertainty, permitting sudden large mass loss consistent with gross fragmentation. At each time step, the relative likelihood of each model is evaluated against the observed positional data, and a smoothed probability is computed over the full trajectory by running the filter both forwards and backwards. Periods where Model 2 dominates (indicated by a switch in smoothed model probability; likelihood $>50\%$) identify where the observed deceleration is inconsistent with steady ablation alone, and gross fragmentation is the more probable explanation. 

For infrasound source attribution, the comparison of different modes is physically meaningful. A fireball trajectory dominated throughout by Model 1 suggests a continuously ablating source, for which we expect a ballistic shock source mechanism. In contrast, trajectories exhibiting clear Model 2 switching periods indicate likely fragmentation events, and the infrasound shock source may be attributed to fragmentation rather than the ballistic trajectory.

\begin{figure}[!h]
    \centering
    \includegraphics[width=0.99\textwidth]{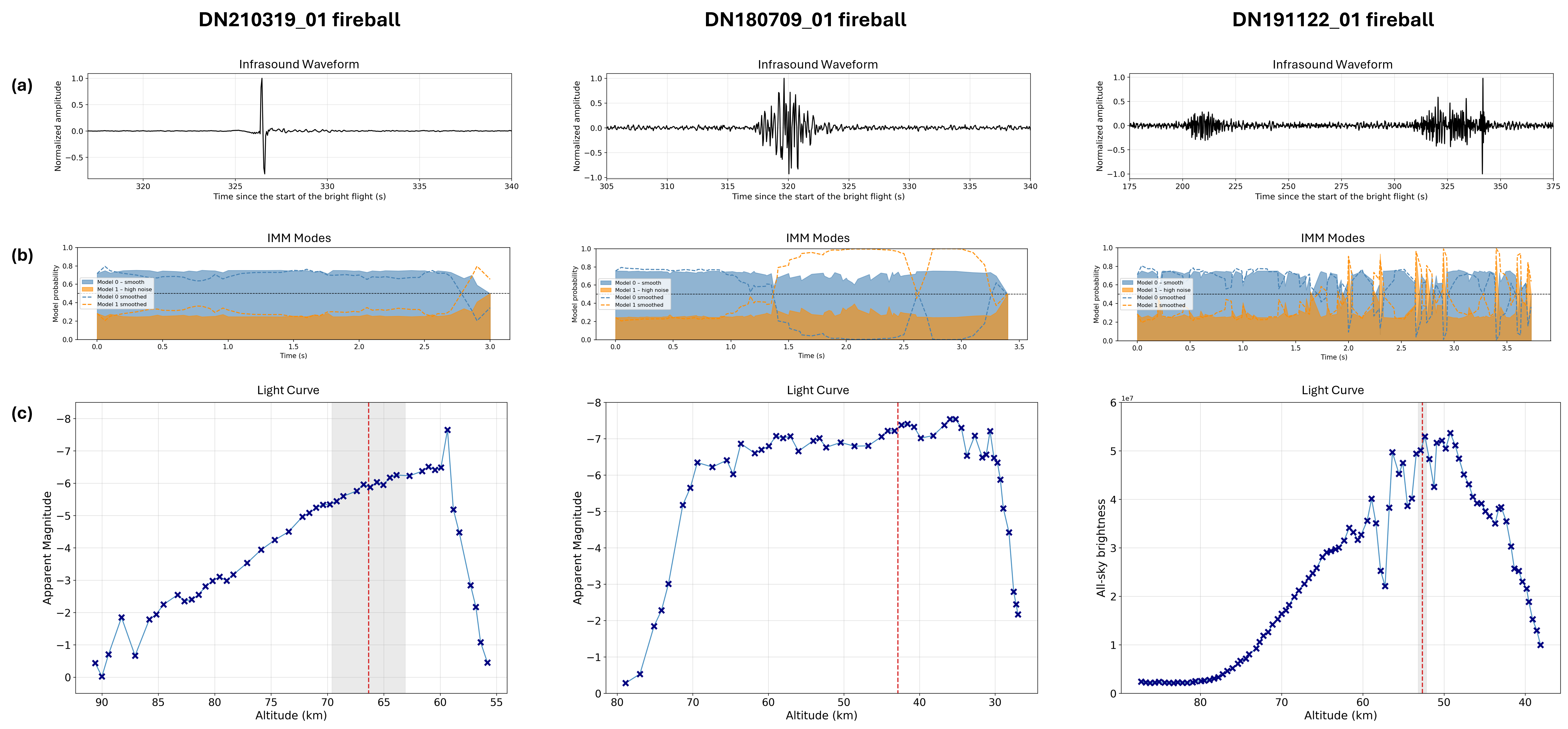}
    \caption{Shock source mechanism analysis for three representative fireball events: DN210319\_01, for which the dominant shock source is likely ballistic, and DN180709\_01 and DN191122\_01, both of which are associated with fragmentation processes. The classification is based on a combination of (a) the recorded infrasound waveforms, (b) the IMM-derived fragmentation behaviour, and (c) the optical light curves. In the light curve panels, the red dashed lines indicate the mean estimated shock source altitude (converted in time for DN191122\_01), while the grey shaded regions represent the associated uncertainty range.}
    \label{fig:shockmechanism}
\end{figure}

Of the 20 events, 10 showed Model 1 to be dominant throughout. All remaining events exhibited a transition to Model 2 becoming dominant ($>50\%$ likelihood) for a period of time (Table \ref{table:IMM}). In some cases, multiple episodes were observed within the range listed in Table \ref{table:IMM}, potentially indicating multiple fragmentation events along the trajectory. In most cases, Model 2 remained dominant throughout the specified duration; however, event DN191122\_01 exhibits eight single-time-step (dt $\sim$ 0.02 s) switches that can be matched to distinct spikes in the fireball's light curve (Fig. \ref{fig:shockmechanism}c (right)).  \\

\begin{table*}[!h]
    \caption{Identified fragmentation episodes for each fireball event, defined as periods where the IMM smoother assigns dominant probability to Model 2 (the high mass-loss covariance model) indicating that observed deceleration is inconsistent with steady ablation alone. Switch onset and offset are given as both altitude (km) and time (s) from the start of the luminous trajectory, defining the altitude window over which fragmentation is probable. Multiple episodes per event could reflect distinct fragmentation periods along the trajectory and N gives the total number of discrete Model 2 episodes identified, where N $>$ 1 indicates the possibility of repeated or episodic fragmentation rather than a single event. $\dagger$ Maximum Model 2 probability does not exceed 60\%, indicating a weak or ambiguous mode switch. * Last observed datapoint along the fireball trajectory.}
    \label{table:IMM}
    \centering
        \begin{tabular}{c| c c |c c |c |c }
        \hline
    \multicolumn{1}{c}{Fireball ID} & \multicolumn{2}{c}{Mode 2 onset}          & \multicolumn{2}{c}{Mode 2 offset}         & \multicolumn{1}{c}{Duration} & \multicolumn{1}{c}{N} \\
    \multicolumn{1}{c}{}            & \multicolumn{1}{c}{alt (km)} & \multicolumn{1}{c}{time (s)} & \multicolumn{1}{c}{alt (km)} & \multicolumn{1}{c}{time (s)} & \multicolumn{1}{c}{(s)}      & \multicolumn{1}{c}{}  \\
            \hline\hline
    DN171009\_01& 54.72    & 1.55     & 27.18    & 3.45     & 1.9      & 2 \\
    DN180709\_01& 52.99    & 1.37     & 27.37    & 3.25     & 1.9      & 2 \\
    DN180828\_01& 79.84    & 0& 59.33    & 1.2      & 1.2      & 2 \\
    DN190824\_02& -& -& -& -& -& 0 \\
    DN191001\_02$^\dagger$ & 60.65    & 1.15     & 31.37    & 2.6      & 1.5      & 1 \\
    DN191002\_01& -& -& -& -& -& 0 \\
    DN191122\_01& 60.16    & 2& 37.83    & 3.73     & 1.7      & 8 \\
    DN200512\_04& 49.03    & 6.2      & 48.64    & 6.3      & 0.1      & 1 \\
    DN200710\_01& -& -& -& -& -& 0 \\
    DN201213\_04& 71.82    & 0.74     & 45.63    & 1.54     & 0.8      & 1 \\
    DN210113\_02& 55.67    & 1.6      & 50.18    & 1.9      & 0.3      & 2 \\
    DN210211\_01& -& -& -& -& -& 0 \\
    DN210319\_01& 57.40    & 2.75     & 55.58    & 3& 0.3      & 1 \\
    DN210418\_02& -& -& -& -& -& 0 \\
    DN211013\_03& -& -& -& -& -& 0 \\
    DN211106\_01& 50.76    & 1.1      & 40.45    & 2.1      & 1.0      & 1 \\
    DN220721\_01& -& -& -& -& -& 0 \\
    DN230629\_02& -& -& -& -& -& 0 \\
    DN231006\_01& -& -& -& -& -& 0 \\
    DN250510\_01& 37.83    & 8.21     & 33.70*   & 9.31*    & 1.1+     & 1+\\
    \hline
        \end{tabular}
\end{table*}

In summary, we applied four independent methods to assess the shock source mechanism of each fireball-generated infrasound signal, classifying events as either ballistic or fragmentation-driven. Ray deviation angles uniquely identified one fragmenting event. The signal morphology analysis suggests an approximately 50:50 split between the cylindrical line-source type signals and the fragmentation-type ones, a result that is broadly supported by the IMM analysis. Light curves usually are a useful indicator of gross fragmentation; however, the resolution of DFN/GFO light curves is not ideal, and many records are affected by saturation. The approach used to determine the final shock source mechanism for each fireball is presented in the \textit{Discussion} section (Table \ref{table:Fragm?}).\\


\section*{Discussion}\label{sec:discussion}

This study investigates infrasound signals from a new high-fidelity dataset of regional fireball events ranging from centimetre- to metre-scale objects. The combination of high-quality CTBTO infrasound data and optically observed trajectories enables assessment of the detection efficiency of small fireballs using an infrasound array, and allows constraints to be placed on the shock source altitudes and mechanisms of the studied fireballs.

\subsection*{Detectability}
Regarding the detection capability of infrasound sensors for small, local-to-regional fireball events, the KDE and PCA analysis results suggest that infrasound-detected fireballs span a broad range of trajectory and physical characteristics (e.g., entry angle, entry velocity, terminal altitude, energy estimates, shock source altitude). This indicates that the detectability of fireball-generated infrasound is likely controlled more strongly by propagation effects (i.e. atmospheric conditions) and source-to-receiver geometry than by any specific fireball parameter. In particular, the lack of clear clustering in parameter space implies that no single physical threshold (e.g., specific velocity range) governs detectability within the analysed dataset. Instead, detection appears to result from a combination of multiple factors rather than a dominant controlling variable. \\

The three events for which detected infrasound arrivals did not match the infraGA simulations (DN161030\_04, DN210904\_01 and DN230302\_02), preventing the determination of their shock source altitudes, highlight the influence of atmospheric conditions on detection. We used a single-line source in this work, which likely contributes to these mismatches, as it may not fully captured the complexity of the propagation paths. This limitation is expected to be more significant for longer-range detections ($>$ 200 km), where propagation is highly sensitive to atmospheric variability and involves ground and/or atmospheric layer reflections. Future works could improve these estimates by using more refined atmospheric profiles, including time-averaged ones, range-dependent propagation simulations mode, or by incorporating radiosonde data.\\

These atmospheric and geometric factors must also be considered together with the shock source mechanism. In particular, cylindrical line source (i.e., ballistic) shock waves are more strongly geometry-dependent than fragmentation-generated signals, since the shock wave is preferentially radiated perpendicular to the fireball trajectory. As a result, the propagation path may not intersect the receiver, reducing the likelihood of detection. We expect this effect to become even more pronounced with increasing source-to-receiver distance. \\

\subsection*{Shock Mechanisms}
Regarding the determination of the shock source mechanism for each fireball, we focused on comparing waveforms characteristics with the IMM analysis results, as the light curve data were not sufficiently reliable for most events, and ray deviation analysis provided a unique solution for only one event (Table \ref{table:Fragm?}).
More specifically, N-wave type arrivals, combined with the absence of fragmentation identification by the IMM and a source altitude corresponding to the closest point along the trajectory, were interpreted as strong evidence for ballistic shock source. In contrast, diffuse waveforms or multiple arrivals, together with IMM-predicted fragmentation altitudes consistent with the estimated source altitude, were considered indicative of a fragmentation shock mechanism. This interpretation is consistent with previous studies of regional meteor-generated infrasound, in which diffuse or multi-arrival waveform morphologies have been associated with fragmentation \citep{silber2014optical, wilson2025bolide, scamfer2026multi}. Fine-scale atmospheric structure can also modify the received waveform by broadening and prolonging the signal, generating diffuse oscillatory tails, and, under some propagation conditions, producing additional arrivals. These effects can complicate attribution of the observed waveform morphology solely to the source process \citep{chunchuzov2025effect, chunchuzov2026multi}.\\

\begin{table*}[!h]
    \caption{Overview of the results obtained from the different methods used to determine the most likely shock source mechanism for each fireball. Columns show the ray deviation angle (degrees), infrasound waveform characteristics, IMM-predicted altitude range (km) over it identifies fragmentation behaviour, estimated source altitude (km), altitude (km) of the closest point (to the sensor) along the fireball trajectory, and the resulting inferred shock mechanism. The horizontal line separates events for which all diagnostic methods are consistent (above) from those exhibiting conflicting results (below).}
    \label{table:Fragm?}
    \centering
    \resizebox{0.99\textwidth}{!}{
        \begin{tabular}{c c c c c c | c c}
        \hline
        Fireball ID & Ray deviation & Waveform & IMM Range & Source Altitude & Altitude Closest Point & Shock Mechanism \\
        \hline
        \hline
DN180709\_01 & 86 & Diffuse & 52.99 - 27.37 & 42.86 & 26.68 & Fragmentation & \multirow{14}{*}{\rotatebox[origin=c]{-90}{\footnotesize{High Confidence}}} \\
DN180828\_01 & 113 & Multiple Arrivals & 79.84 - 59.33 & $79.65\pm0.25$ & 52.97 & Fragmentation \\
DN190824\_02 & 82 & N-wave & - & $52.64\pm7.25$ & 44.77 & Ballistic \\
DN191002\_01 & 66 & N-wave & - & $48.21\pm10.25$ & 37.83 & Ballistic \\
DN191122\_01 & 88 & Multiple Arrivals & 60.16 - 37.83 & $53.67\pm0.50$ & 38.06 & Fragmentation \\
DN200710\_01 & 88 & N-wave & - & $87.12\pm0.25$ & 89.37 & Ballistic \\
DN201213\_04 & 57 & Diffuse & 71.82 - 45.63 & $49.38\pm6.50$ & 42.87 & Fragmentation \\
DN210113\_02 & 90 & Multiple Arrivals & 55.67 - 50.18 & $56.52\pm4.25$ & 46.27 & Fragmentation \\
DN210211\_01 & 110 & N-wave & - & 72.27 & 67.73 & Ballistic \\
DN211013\_03 & 80 & N-wave & - & $41.12\pm2.25$ & 32.36 & Ballistic \\
DN211106\_01 & 85 & Diffuse & 50.76 - 40.45 & $52.39\pm1.75$ & 36.28 & Fragmentation \\
DN220721\_01 & 91 & N-wave & - & $48.77\pm1.50$ & 49.17 & Ballistic \\
DN230629\_02 & 95 & N-wave & - & $60.41\pm4.50$ & 64.91 & Ballistic \\
DN250510\_01 & - & Diffuse & 37.83 - 33.70 & - & 33.70 & Fragmentation \\
\hline
DN171009\_01 & 79 & Multiple Arrivals & 54.72 - 27.18 & $80.23\pm3.5$ & 58.94 & Fragmentation & \multirow{6}{*}{\rotatebox[origin=c]{-90}{\footnotesize{Lower Confidence}}} \\
DN191001\_02 & 91 & N-wave & 60.65 - 31.37 & $59.38\pm15.50$ & 54.76 & Ballistic \\
DN200512\_04 & 94 & N-wave & 49.03 - 48.64 & $71.86\pm0.50$ & 72.08 & Ballistic \\
DN210319\_01 & 92 & N-wave & 57.40 - 55.58 & $66.35\pm3.25$ & 68.17 & Ballistic \\
DN210418\_02 & 100 & Diffuse & - & $64.83\pm0.50$ & 42.50 & Fragmentation \\
DN231006\_01 & 88 & Multiple Arrivals & - & $59.15\pm0.75$ & 50.74 & Fragmentation \\
        \hline
        \end{tabular}
    }
\end{table*}

Among the 20 fireball events, 14 show complete agreement between the waveform characteristics and IMM predictions (all events listed above the horizontal line in Table \ref{table:Fragm?}). Of these, seven are confidently interpreted as ballistic shock sources, including four for which the estimated source altitude coincides with the altitude of the closest point (to the receiver) along the trajectory. The remaining seven fireballs are associated with fragmentation shock sources; and for six of these, the IMM-predicted altitude range is consistent with the estimated source altitude. For the seventh event (DN250510\_01), the source altitude could not be determined, preventing further assessment of altitude agreement. \\

On the other hand, six fireballs in the dataset show conflicting results (events listed below the horizontal line in Table \ref{table:Fragm?}). These cases are discussed below. 

\noindent One event (DN171009\_01) exhibits a signal with multiple arrivals (typically interpreted as evidence of fragmentation) and is also classified as fragmenting by the IMM analysis. However, its estimated source altitude is consistent with neither the IMM altitude range nor the altitude of the closest point, likely due to atmospheric effects. Hence, based on both the waveform characteristics and the IMM classification, we interpret this event as being fragmentation-driven. 

\noindent Three events with suspected ballistic origin (N-wave signatures) exhibit IMM-predicted fragmentation (DN191001\_02, DN200512\_04 and DN210319\_01). 
For all three, their estimated source altitudes are consistent with the altitude of the closest point along the trajectory. 
Despite DN191001\_02 also being consistent with the IMM altitude range, we favour a ballistic shock source origin for all three based on the waveform characteristics and the likely propagation from the shortest distance to the trajectory. 

\noindent Two additional events (DN210418\_02 and DN231006\_01) display waveforms characteristics indicative of fragmentation (either diffuse signal or multiple arrivals), while the IMM does not predict fragmentation in these cases. In addition, their estimated shock source altitudes do not coincide with the altitude of the closest point. Therefore, based on the waveform evidence, we also classify these events as fragmentation-driven. \\

The DN210319\_01 fireball represents a particularly interesting case within our dataset and warrants a more detailed discussion. The recorded infrasound signal exhibits a clear N-wave with a high signal-to-noise ratio (see Fig. \ref{fig:shockmechanism}a), a characteristic typically associated with ballistic shock sources. However, the IMM analysis predicts fragmentation beginning at 2.75 s after entry (Fig. \ref{fig:shockmechanism}b), corresponding to an altitude of approximately 57.4 km. This fragmentation episode also appears in the optical data, with the light curve displaying a spike at an altitude of approximately 59 km (Fig. \ref{fig:shockmechanism}c). 

In contrast, the estimated shock source altitude for this event ranges between 63.1 to 69.6 km, well above the IMM-identified fragmentation altitude. This interval however includes the altitude of the closest point along the trajectory (68.2 km). Therefore, although the IMM analysis appears to successfully identify a late-stage fragmentation episode, both the estimated source altitude and the waveform characteristics favour a ballistic origin for the recorded infrasound signal.

Thus, this event demonstrates the importance of combining multiple observational approaches when investigating fireball source shock mechanisms. One possible explanation in this case, is that the fragmentation-generated acoustic signal was quite weak and/or the event did not present a favourable source-receiver geometry for detection. Consequently, fragmentation processes identified in optical data may not always be detected by infrasound, emphasizing one more time that detectability depends not only on sources characteristics but also on propagation and geometry effects.

\subsection*{Fireball Classification}
\begin{figure}[!h]
    \centering
    \includegraphics[width=0.8\textwidth]{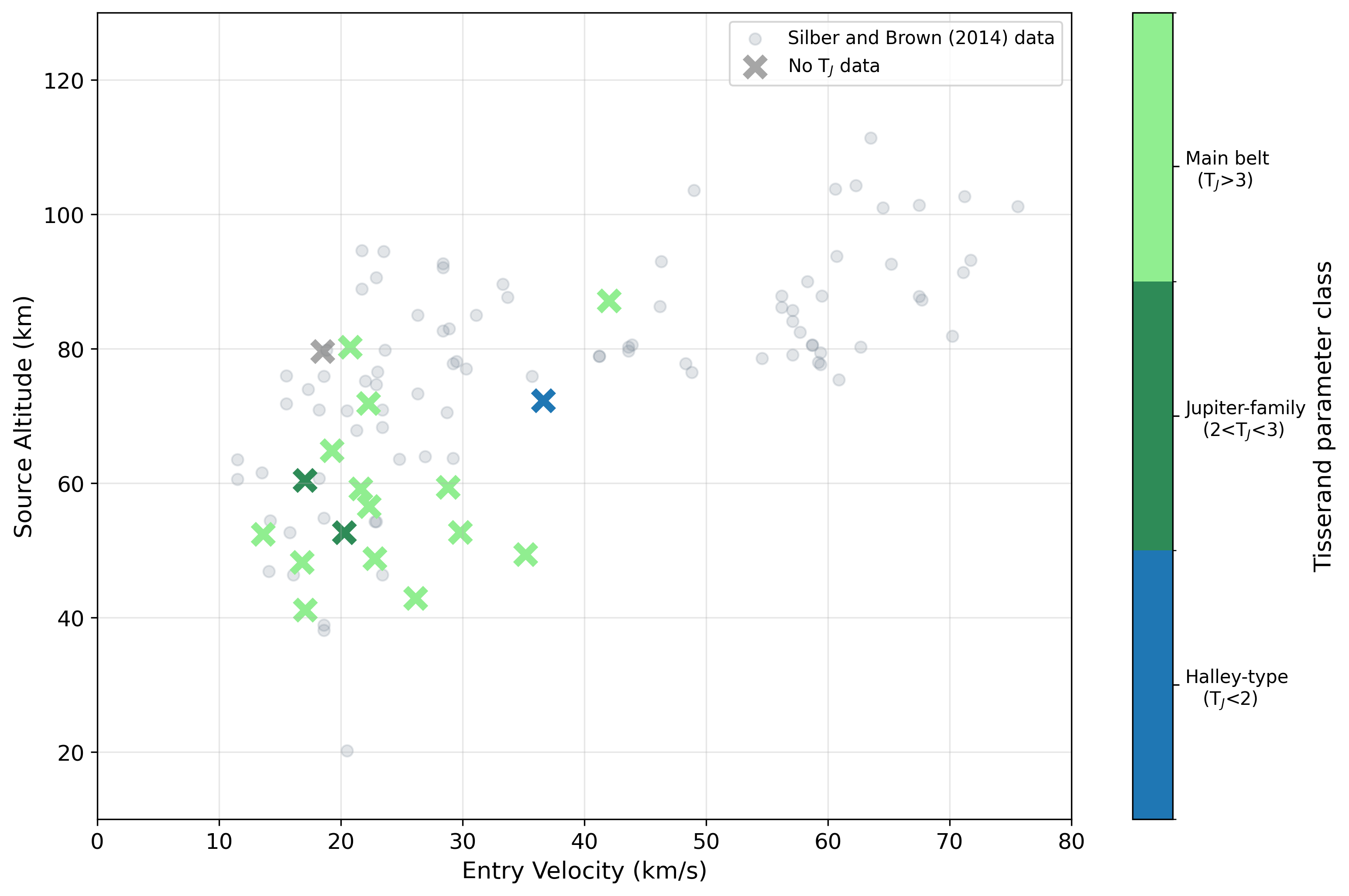}
    \caption{Fireball entry velocity (km/s) as a function of the estimated shock source altitude (km) for the events analysed in this study (crosses), compared with the dataset of \citet{silber2014optical} (circles). The colour of the crosses indicates the orbital family associated with each fireball, based on the Tisserand parameter classification (i.e., Main belt, Jupiter-family, or Halley-type orbits).}
    \label{fig:tj}
\end{figure}

\begin{figure}[!h]
    \centering
    \includegraphics[width=0.9\textwidth]{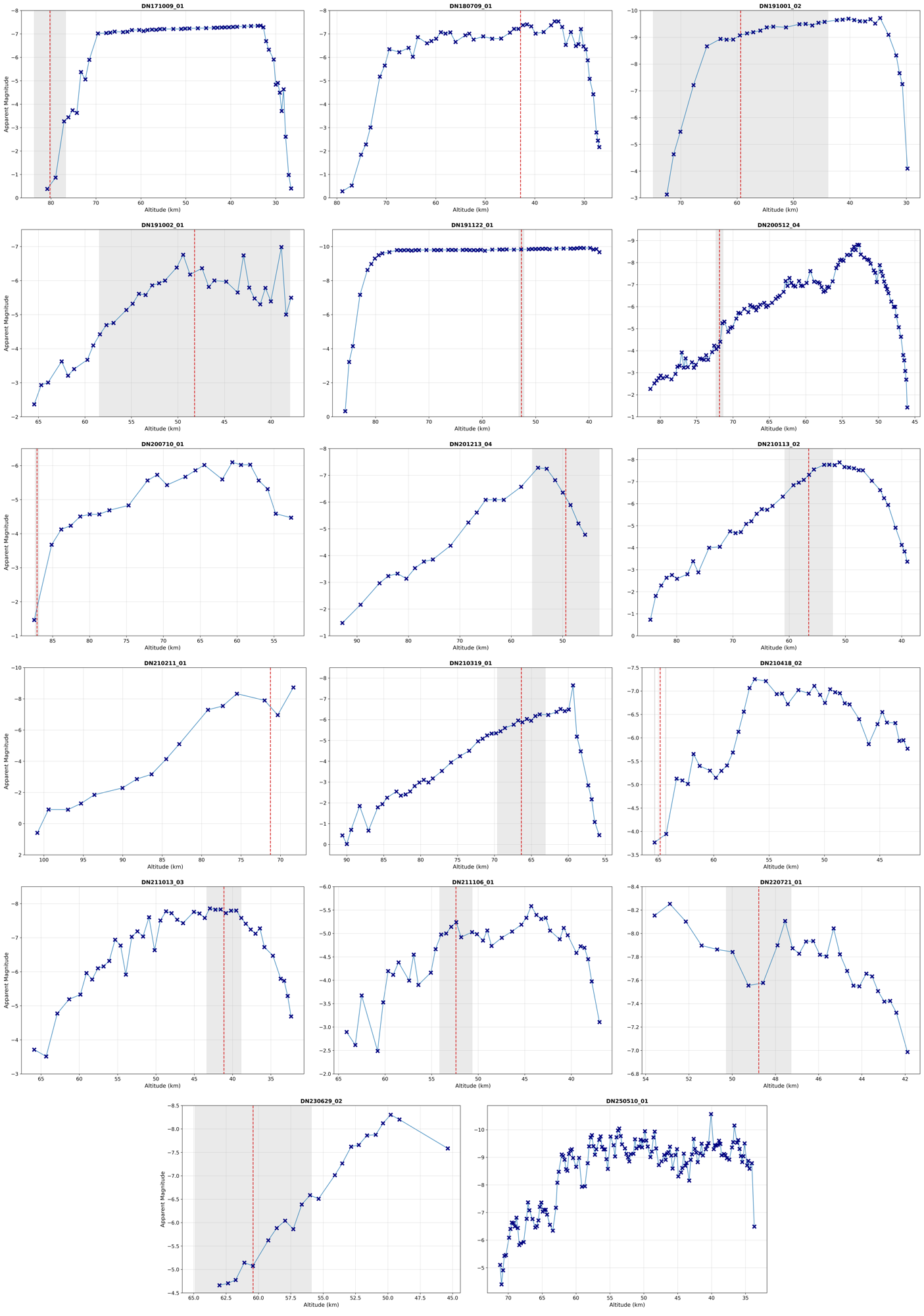}
    \caption{Compilation of light curves for fireball events with confirmed infrasound detections. The red dashed line indicates the mean source altitude, while the grey shaded area represents the corresponding altitude uncertainty range. Light curves exhibiting a pronounced flat top are classified as saturated.}
    \label{fig:LCs}
\end{figure}

For events for which a source altitude could be determined, we observe a relationship between fireball entry velocity and shock source altitude similar to that reported by \citet{silber2014optical} (see Fig. \ref{fig:tj}). In their study, fireballs were divided into two populations based on entry speed: a low-velocity population ($<$ 40 km/s) and a high-velocity population ($>$ 40 km/s). These two groups exhibit distinct characteristics, including differences in inferred meteoroid origin (i.e., asteroidal orbit versus cometary orbit), luminous trail duration, and final altitude. 
Most events in our dataset appear to be consistent with the low-velocity population according to their estimated source altitudes and entry velocities, suggesting predominantly asteroidal-type material. This is also consistent with the typical size range of DFN/GFO detections (cm- to dm-scale objects), as cometary material events are generally weaker and less likely to survive deep into the atmosphere. \\

However, two events display characteristics inconsistent with their source altitude-entry velocity classification. Event DN210211\_01 falls within the low-velocity population, yet several of its properties are more consistent with the high-velocity family. In particular, its orbit indicates a Halley-type cometary origin (Tisserand parameter of 1.05), exhibits a short duration luminous flight ($\approx$1 s), and reached a relatively high final altitude ($\approx$68 km), exceeding the 63 km threshold identified by \citet{silber2014optical}. Conversely, event DN200710\_01 falls within the high-velocity population based on its entry speed, but several of its other characteristics resemble those of the low-velocity family, including orbital properties suggesting a main belt origin. 
The apparent ‘misclassification’ of DN210211\_01 may reflect the small but non-negligible contamination of Jupiter-family comet-like orbits by cometary-origin objects, as observed by \citet{shober2021main}. \\

Meteoroid mass estimation is beyond the scope of the present study and is reserved for \textit{Part II}. The follow-up analysis will use the same fireball dataset to estimate meteoroid masses from the infrasound observations and compare them with independent mass estimates obtained using established methods.\\

\subsection*{Relevance to Planetary Defense and Atmospheric Re-entry Monitoring}

Small fireballs provide frequent opportunities to study the energy deposition, fragmentation, and atmospheric shocks associated with natural entries, improving our ability to interpret larger events of greater concern to planetary defense. In this study, infrasound complements optical observations by providing independent estimates of event energy and acoustic source altitude and by helping identify the processes responsible for shock generation. These capabilities are also becoming increasingly relevant as growing space traffic leads to more spacecraft returns, including Orion missions, and more uncontrolled space-debris reentries. The well-documented entries of Hayabusa2 and OSIRIS-REx have enabled direct tests of seismoacoustic detection, propagation, and source-reconstruction methods \citep{clemente2025exploring, silber2024geophysical, fernando2025array, nishikawa2026optimizing, scamfer2026multi}, while the OSIRIS-REx measurements have also been used to evaluate cylindrical weak-shock theory against known capsule parameters \citep{silber2026benchmarking}. For less well-constrained events, ground-based observations can recover information that is no longer available from orbital tracking, as demonstrated by the reconstruction of the trajectory and breakup of an uncontrolled Soyuz-2.1b reentry \citep{hatty2026tracking}.

\section*{Conclusions}

In this paper, we analysed fireball-generated infrasound signals associated with events for which high-fidelity optical trajectories are available. The objectives of this study were to: (i) estimate entry dynamics of fireballs and assess the detection capability of infrasound sensors for small, regional events; and (ii) identify the shock source mechanisms responsible for the recorded infrasonic signals (i.e., ballistic versus fragmentation sources). \\

(i) Results from the KDE analysis indicate that fireballs detected both optically and infrasonically span a broad range of dynamical parameters, with detected events showing no preference or trend in comparison to non-detected fireball events. This suggests that no unique set of entry conditions governs detectability. This is further reflected in the broad range of estimated energies and shock source altitudes of the detected events ($1.719 \times 10^{-5}$--$1.799 \times 10^{-2}$ kt TNT eq. and 41--87 km, respectively). Furthermore, inferred shock source locations occur throughout the fireball trajectories rather than being confined to a particular segment. Together, these observations suggest that fireball infrasound detectability is not controlled by a single dominant parameter, but instead reflects the combined influence of atmospheric propagation conditions and source-receiver geometry.\\

(ii) Results from the shock source mechanisms analysis indicate an even split between ballistic and fragmentation-related sources among the studied fireballs, highlighting the diversity of processes responsible for observed infrasound signal. However, the DN210319\_01 event case illustrates the limitations of relying on a single diagnostic method to determine whether a meteoroid fragmented and to constrain the shock source mechanism. While the recorded infrasound signal exhibits characteristics consistent with a ballistic shock source, the light curve and IMM analyses indicate a late-stage fragmentation episode. This fragmentation does not appear to have generated a detectable infrasound signature, suggesting that ballistic shocks may dominate the recorded signal when fragmentation contributions are weak, not efficiently coupled into the atmosphere, or with a non-favourable source-receiver geometry.

\noindent This case demonstrates that fragmentation events, although often expected to produce strong infrasonic signals, do not necessarily dominate the observed waveforms. In contrast, the detectability of a given shock source appears to depend more strongly on atmospheric propagation conditions and source-receiver geometry factors. Consequently, these results suggest that infrasound observations alone may not fully capture the complexity of fireball-generated signals, but highlight the value of combining optical and infrasonic observations to better constrain shock source mechanisms.\\

\section*{Acknowledgments}
The authors acknowledge the use of data from the International Monitoring System (IMS) operated by the Comprehensive Nuclear-Test-Ban Treaty Organization (CTBTO). Details on the CTBTO data used in this study are provided in the \textit{Data Availability} section. The Desert Fireball Network and Global Fireball Observatory programs have been funded by the Australian Research Council as part of the Australian Discovery Project scheme (DP170102529, DP200102073, DP230100301), the Linkage Infrastructure, Equipment and Facilities scheme (LE170100106), and receives institutional support from Curtin University. Geoscience Australia is also gratefully acknowledged for hosting a research internship and for valuable guidance in data acquisition and processing. The authors would also like to thank the hosts of our DFN cameras, both in Australia and in the United States of America. EAS was supported by the Laboratory Directed Research and Development (LDRD) program (project number 229346 and 236896) at Sandia National Laboratories, a multimission laboratory managed and operated by National Technology and Engineering Solutions of Sandia, LLC., a wholly owned subsidiary of Honeywell International, Inc., for the U.S. Department of Energy’s National Nuclear Security Administration under contract DE-NA0003525. 
This article has been authored by an employee of National Technology \& Engineering Solutions of Sandia, LLC under Contract No. DE-NA0003525 with the U.S. Department of Energy (DOE). The employee owns all right, title and interest in and to the article and is solely responsible for its contents. The United States Government retains and the publisher, by accepting the article for publication, acknowledges that the United States Government retains a non-exclusive, paid-up, irrevocable, world-wide license to publish or reproduce the published form of this article or allow others to do so, for United States Government purposes. The DOE will provide public access to these results of federally sponsored research in accordance with the DOE Public Access Plan \url{https://www.energy.gov/downloads/doe-public-access-plan}. This paper describes objective technical results and analysis. Any subjective views or opinions that might be expressed in the paper do not necessarily represent the views of the U.S. Department of Energy or the United States Government.

\section*{Data Availability}
The data (infrasound) used in this study are available directly from CTBTO virtual Data Exploitation Centre (vDEC). Data access can be granted upon request, following signature of a contract after necessary approvals have been secured. The views expressed in this paper are those of the authors and do not necessarily reflect those of the CTBTO Preparatory Commission. The trajectory data for the studied events are provided in Table \ref{table:GeneralInfoFB} (Appendix).

\section*{Declaration of competing interests}
The authors acknowledge no conflicts of interest.

\bibliography{Refs}

\clearpage

\clearpage
\section*{Appendix}

\begin{table*}[!h]
    \caption{Summary of the studied fireball events, including timing, duration, initial velocity, and trajectory endpoints.}
    \label{table:GeneralInfoFB}
    \centering
    \resizebox{0.99\textwidth}{!}{
        \begin{tabular}{lcccccccccc}
            \hline
            Fireball ID & Datetime & Duration & Initial Velocity  & Initial Height  & Initial Latitude & Initial Longitude & Final Height  & Final Latitude & Final Longitude & Shortest 3D Distance \\
             &  (UTC) &  (s) &  (m/s) &  (m) &  &  & (m) &  &  &  (km)\\
            \hline\hline
            \midrule
            DN161030\_04 & 2016-10-30T14:47:07.400 & 5.35  & 16602 & 72794 & -33.7933 & 120.2821 & 32024 & -34.3700 & 120.3627 & 367.3\\
            DN171009\_01 & 2017-10-09T17:40:22.250 & 3.65  & 20770 & 83727 & -33.5357 & 117.3516 & 26072 & -33.2907 & 117.3451 & 167.4\\
            DN180709\_01 & 2018-07-09T20:16:19.900 & 3.4   & 26141 & 78862 & -32.8302 & 115.4105 & 26675 & -32.8792 & 115.9333 & 198.0\\
            DN180828\_01 & 2018-08-28T11:40:47.700 & 1.6   & -     & 79901 & -31.7253 & 117.2587 & 52974 & -31.7973 & 117.3136 & 327.3\\
            DN190824\_02 & 2019-08-24T07:07:23.476 & 1.524 & 20287 & 72385 & 33.7866  & -115.1492 & 44773 & 33.8268  & -115.0610 & 142.7\\
            DN191001\_02 & 2019-10-01T09:27:10.750 & 2.75  & 28811 & 87879 & 33.9898  & -116.0465 & 29798 & 34.3076  & -115.8492 & 98.7\\
            DN191002\_01 & 2019-10-02T16:20:18.300 & 2.1   & 16823 & 65457 & -34.3975 & 116.7679 & 37831 & -34.2341 & 116.7795 & 65.6\\
            DN191122\_01 & 2019-11-22T14:15:51.750 & 3.7   & 29875 & 87761 & -32.1950 & 120.3374 & 38058 & -31.7145 & 119.4991 & 433.1\\
            DN200512\_04 & 2020-05-12T21:53:17.500 & 7.1   & 22274 & 81360 & -34.4152 & 116.1404 & 46052 & -34.0027 & 117.5346 & 79.4\\
            DN200710\_01 & 2020-07-10T20:57:22.400 & 1.5   & 42039 & 89369 & -33.5783 & 115.5810 & 52687 & -33.4520 & 115.0867 & 163.0\\
            DN201213\_04 & 2020-12-13T10:56:51.400 & 1.7   & 35176 & 96885 & 34.2496  & -116.3339 & 42870 & 34.2603  & -116.1932 & 89.7\\
            DN210113\_02 & 2021-01-13T04:50:11.000 & 2.7   & 22327 & 85771 & 34.6097  & -117.2528 & 39016 & 34.5719  & -117.5619 & 150.6\\
            DN210211\_01 & 2021-02-11T16:09:16.350 & 1     & 36636 & 102766 & -31.5847 & 116.1409 & 67732 & -31.6634 & 116.1020 & 334.6\\
            DN210319\_01 & 2021-03-19T20:59:13.900 & 3     & -     & 90598 & -34.9236 & 116.0663 & 55812 & -35.1595 & 116.8252 & 87.4\\
            DN210418\_02 & 2021-04-18T12:03:42.200 & 2.45  & 19279 & 65333 & -31.8754 & 113.3247 & 42504 & -31.6627 & 113.6326 & 418.2\\
            DN210904\_01 & 2021-09-04T06:47:42.171 & 2.129 & 17253 & 60092 & 35.4002  & -114.5755 & 30333 & 35.5523  & -114.5670 & 273.4\\
            DN211013\_03 & 2021-10-13T06:14:00.400 & 2.55  & 17079 & 67372 & 33.0377  & -117.6648 & 32363 & 33.1973  & -117.6228 & 118.0\\
            DN211106\_01 & 2021-11-06T14:38:16.500 & 2.6   & 13624 & 64136 & -33.5293 & 116.6136 & 36280 & -33.5131 & 116.7957 & 131.3\\
            DN220721\_01 & 2022-07-21T12:02:00.043 & 1.857 & 22792 & 53769 & -34.1352 & 115.9233 & 41892 & -34.3336 & 115.6171 & 85.2\\
            DN230302\_02 & 2023-03-02T10:20:54.757 & 1.039 & 47680 & 101015 & 34.8177  & -121.5164 & 61094 & 34.7099  & -121.9315 & 491.2\\
            DN230629\_02 & 2023-06-29T11:22:38.250 & 2.15  & 17057 & 64912 & -34.3247 & 115.3393 & 39839 & -34.4195 & 115.1191 & 120.9\\
            DN231006\_01 & 2023-10-06T10:04:03.255 & 2.55  & 21547 & 70951 & 34.4210  & -116.0887 & 34780 & 34.4173  & -115.7252 & 118.5\\
            DN250510\_01 & 2025-05-10T21:57:42.193 & 9.307 & 14757 & 71999 & -31.1596 & 119.7835 & 33701 & -32.2422 & 120.2336 & 443.2\\
            \bottomrule
            \hline
        \end{tabular}
        }
\end{table*}

\end{document}